\documentclass[12pt]{article}
\usepackage{amsmath, amssymb, graphics, graphicx}

\newcommand{\sect}[1]{\setcounter{equation}{0}\section{#1}}

\def\be{\begin{equation}}
\def\ee{\end{equation}}
\def\ba{\begin{eqnarray}}
\def\ea{\end{eqnarray}}

\title{Perturbations around the near horizon limit of charged Randall Sundrum black holes}
\author{Alexander Kaus \\ \\ Department of Applied Mathematics and Theoretical Physics\\ Centre for Mathematical Sciences \\ Wilberforce Road, Cambridge CB3 0WA, UK \\ ak417@cam.ac.uk}
\begin{document}

\maketitle

\begin{abstract}
In a previous paper we determined the near horizon limit of an extremal brane world black hole, charged with respect to a Maxwell field on the brane, in the single brane Randall-Sundrum model. This paper is largely an extension of that work. The same black hole is considered. A metric expansion around the near horizon limit is set up and the correction terms of the first two subleading orders are determined. It is found that the corrected bulk metric can still be sliced by a brane even though the $Ads_{2}$ symmetry of the near horizon metric is broken by correction terms. The induced metric on the brane is determined to second correction order and compared with the predictions of 4d General Relativity. It is found that large black holes asymptote the extremal Reissner Nordstr\"om solution and thus agreement with 4d General Realtivity is obtained in this limit. 
\end{abstract}

\sect{Introduction}

Our best candidate for a theory of Quantum Gravity is String Theory. However for String Theory to be consistent, spacetime has to be higher-dimensional. For some time the only consistent way to resolve the apparent contradiction between theory and experiment was to assume that the extra dimensions are compactified on an internal manifold whose size is too small to be observed by experiment. However the realisation that String Theory naturally gives rise to branes led Randall and Sundrum (RS) to discover that non-compact extra dimensions not neccessarily contradict observation \cite{RS1,RS2}. In their model they consider our universe being a 4 dimensional brane in a 5 dimensional bulk spacetime with negative cosmological constant. They prove that on the perturbative level the theory induced on the brane behaves like Newtonian Gravity as long as distances are long compared to the $AdS$ length \cite{RS2, perturb}. This led to the study of brane world gravity and brane world black holes in particular \cite{Maartens:2003tw,gregory}, since it was unclear whether agreement between the induced theory on the brane and 4d Gravity extended beyond perturbation theory.
	 
	 Although there was a lot of effort to find brane world black hole solutions none could be constructed analytically. Even numerically it has so far been impossible to find a satisfactory answer although some progress was made \cite{numerical}. The only case in which a full analytic solution could be found was in one lower dimension. By considering the C-metric Emparan, Horowitz and Myers were able to construct brane world black holes on flat branes \cite{ehm1} and branes with negative cosmological constant \cite{ehm2} in the lower dimensional equivalent of the Randall-Sundrum model. However in contrast to the perturbative results of Randall and Sundrum where agreement with $4d$ GR was observed, in the lower dimensional case the induced theory on the brane significantly differs from $3d$ GR. 
	 
	 A possible explanation for the impossibility to find brane world black hole solutions in the RS model was given by work based on the AdS$/$CFT correspondence \cite{maldacena}. According to work on that correspondence \cite{gubser} the RS model is equivalent to a four dimensional effective theory of General Relativity coupled to a cut-off CFT. Thus a brane world black hole in the RS model would be equivalent to a quantum-corrected black hole \cite{efk}. That however means that one expects the black hole on the brane to Hawking radiate and thus not to be static \cite{efk, tanaka}. This would of course explain why it has been impossible to obtain any kind of static solution. However it has also been pointed out that the validity of this argument may be flawed due to strong coupling effects, since these may lead to Hawking radiation being further suppressed than one would expect naively \cite{wiseman}.
	 
	 One way to avoid the issue of whether a static solution exists or not is to consider extremal black holes. Such black holes have zero temperature and thus would not Hawking radiate in the dual theory. In a previous paper \cite{Kaus:2009cg} we thus considered extremal, static black holes which are charged with respect to a Maxwell field on the brane and which are spherically symmetric on the brane. The further advantage of considering extremal black holes is that one can always solve for their near-horizon geometry instead of their full solution \cite{Reall:2002bh}, which is a significantly easier problem. A main reason for doing this is that the black hole entropy can be determined from the near horizon limit alone and thus a meaningful comparison to $4d$ GR is still possible. In \cite{Kaus:2009cg} the near-horizon limit of the brane world black hole was determined and for the induced theory on the brane agreement with $4d$ GR was observed in the limit of large black holes. 
	 
	 The main disadvantage of only knowing the near-horizon limit is that no statements can be made concerning whether it can be extended to a full brane world black hole solution. This paper partially answers that question. In the coming chapters a metric expansion around the near horizon limit of the bulk metric is set up and the first two subleading orders are determined. Partially this is done analytically and partially by relying on numerical methods. Having determined the bulk solution to second correction order the corresponding Israel matching conditions are calculated and it is investigated whether these can be fulfilled by the bulk metric including corrections. In \cite{klr} it was proven that there is a symmetry enhancement of the near horizon geometry of extremal black holes. It always contains a two dimensional maximally symmetric spacetime as a subspace. In \cite{Kaus:2009cg} it was shown that in the case of the black hole in question this 2 dimensional space is $AdS_2$. It was this symmetry enhancement that made it mathematically clear why one would expect the near horizon limit of the bulk metric to allow for a brane slicing. However this $AdS_2$ symmetry is broken by correction terms when we go beyond the near horizon limit. Nonetheless it is found that the bulk can still be sliced by a brane. This is notable since due to the breaking of the $AdS_2$ symmetry there are potentially more independent constraint equations than free parameters at each order. The induced metric on the brane is determined and compared to the relevant $4d$ GR metric. Good agreement between the two is observed for black holes large compared to the AdS scale. Calculations beyond the second order correction terms which are not presented expicitly in this paper seem to indicate that the findings for the second order terms are stereotypical for any correction order $\geq 2$. Even though this paper does not prove that the near horizon limit can be extended to a full brane world black hole solution, it gives strong evidence supporting that case.

\sect{Bulk}
A particular RS II setup is considered which consists of an extremal static charged brane world black hole. It is natural to assume that the extension of the brane world black hole to the bulk is static as well. Furthermore, since the surface gravity is constant in the bulk, it follows by continuity that it must take the same value in the bulk as it does on the brane. Thus if the horizon is degenerate on the brane it will also be degenerate in the bulk. Following \cite{klr} and introducing Gaussian null coordinates $(v,r,x^{a})$ in the neighbourhood of the horizon, the bulk metric can be put in the form

	\be
	ds^2=K(x,r)dv^2 +2dv dr+ L_{a}(x,r) dv dx^{a}+h_{ab}(r,x)dx^{a}dx^{b} .
 \ee

$\frac{\partial}{\partial v}$ generates time translations. The horizon is located at $r=0$ and $K(x,r)=O(r^2)$ due to extremality. We are only interested in the case of a spherically symmetric brane world black hole. This allows us to introduce coordinates $x^{a}=(\rho,\theta,\phi)$ making the bulk metric axisymmetric and simplyfying it to
	\be
	ds^2=K(\rho,r)dv^2 +2dv dr + L(\rho,r) dv d\rho + M(\rho,r)^2 d\rho^2 + N(\rho,r)^2 d\Omega^2.
\ee
Taking the near horizon limit of the above metric, defined by 
	\be
	r \rightarrow \epsilon r \qquad v \rightarrow \frac{v}{\epsilon} \qquad \epsilon \rightarrow 0 ,
\ee
simplifies it to
	\be
	ds^2=\tilde{K}(\rho)r^2dv^2 +2dv dr + \tilde{L}(\rho)r dv d\rho + \tilde{M}(\rho)^2 d\rho^2 + \tilde{N}(\rho)^2 d\Omega^2.
\ee
In \cite{klr} it was proven that there is a symmetry enhancement of the near horizon metric. Following the methods applied in \cite{klr} and changing gauge to a coordinate system in which the extra symmetry is more evident, the above metric becomes
	\be
	ds^2=\hat{K}(\rho)\Big(k\hat{r}^2dv^2 +2dv d\hat{r} \Big) + \hat{M}(\rho)^2 d\rho^2 + \hat{N}(\rho)^2 d\Omega^2,
\ee
where $k$ can take the values $-1$, $0$ or $1$. In \cite{Kaus:2009cg}  
it was determined that $k=-1$ for the brane world black hole considered here. Using the remaining gauge freedom associated with the cooradinate $\rho$ we can furthermore set $\hat{M}(\rho)=1$. Renaming metric functions and coordinates then leads to the near horizon metric determined in \cite{Kaus:2009cg} given by
\be
\label{nearhorizon}
	ds^2= A(\rho)^2 d\Sigma^2 + d\rho^2 + R(\rho)^2 d\Omega^2 ,
\ee
where $d\Sigma^2$ is the metric on $AdS_{2}$ of unit radius. The equations governing $A(\rho)$ and $R(\rho)$ were determined to be
	\be
	\frac{1}{A^2}+\frac{A'^2}{A^2}+\frac{2A'R'}{AR}+\frac{A''}{A}=\frac{4}{l^2} ,
	\label{Aeom}
\ee
	\be
		\frac{-1}{R^2}+\frac{R'^2}{R^2}+\frac{2A'R'}{AR}+\frac{R''}{R}=\frac{4}{l^2} ,
		\label{Reom}
\ee
where $l$ is the AdS radius of curvature. These were solved numerically in \cite{Kaus:2009cg}. Furthermore in that paper a Taylor series expansion of those functions about $\rho=0$ was determined, whose first terms are given by
	\be
	A = A_{0}+\Big(\frac{-1}{A_{0}}+\frac{4 A_{0}}{l^2}\Big)\frac{\rho^2}{6}+\Big(-\frac{11}{ A_{0}^3}+\frac{40}{ A_{0}l^2}+\frac{16 A_{0}}{l^4}\Big)\frac{\rho^4}{1080}+\ldots
	\label{Ataylor}
\ee
	\be
	R = \rho+\Big(\frac{1}{ A_{0}^2}+\frac{2}{l^2}\Big)\frac{\rho^3}{18}+\Big(\frac{53}{ A_{0}^4}-\frac{220}{l^2 A_{0}^2}+\frac{212}{l^4}\Big)\frac{\rho^5}{5400}+\ldots
	\label{Rtaylor}
\ee
i.e. the set of solutions forms a 1-parameter family labelled by the free parameter $A_{0}$.

	With $A(\rho)$ of the near horizon metric given numerically, we change the gauge of the general metric slightly by $r \rightarrow r A(\rho)^2$ so that the new Ansatz for the bulk metric is of the form
	\be
	\label{generalmetric}
	ds^2= A(\rho)^2(F(\rho,r)dv^2 +2dv dr) + G(\rho,r) dv d\rho + H(\rho,r)^2 d\rho^2 + I(\rho,r)^2 d\Omega^2.
\ee
The horizon is still located at $r=0$ and we still have $F(\rho,r)=O(r^2)$ due to extremality. The above gauge will be used for the remainder of this paper. In the following subsections the four undetermined metric functions will be determined as a power series expansion in $r$ around $r=0$. 
\subsection{First order corrections}
Using the zeroth order results from (\ref{nearhorizon}) we have to first correction order
	\[
	F(\rho,r)=-r^2+f_{1}(\rho)r^3+O(r^4), \qquad G(\rho,r)=0\cdot r+g_{1}(\rho)r^2+O(r^3),
\]
	\be
	H(\rho,r)=1+h_{1}(\rho)r+O(r^2), \qquad I(\rho,r)=R(\rho)+r_{1}(\rho)r+O(r^2),
	\label{firstorder}
\ee
where $f_{1}(\rho)$, $g_{1}(\rho)$, $h_{1}(\rho)$ and $r_{1}(\rho)$ are to be determined. The equations governing these are determined by series expanding the Einstein equations as a power series in $r$ around $r=0$. Requiring the first order corrections in this series to vanish gives rise to the equations of motion of the above functions. These equations are given by
	
\begin{align}
	3A f_{1} R^2-A h_{1} R^2-2A R r_{1} +2g_{1} R^2 A' +2A h_{1} R^2 A'^2+A R^2 g_{1}' +A^2 R^2 A' h_{1}' \notag
	\\ + 2 A g_{1}R R' +4 A^2 h_{1} R A' R' +2 A^2 r_{1} A' R' - 2 A^2 R A' r_{1}' + 2 A^2 h_{1} R^2 A''=0,
\label{firstordereomstart}
\end{align}
\begin{align}
	g_{1}R+A h_{1}R A' +2A r_{1}A' +2 A^2 h_{1} R' -2 A^2 r_{1}'=0,
\end{align}
	
\begin{align}
	l^2h_{1}R^2-4A^2h_{1}R^2-l^2R^2g_{1}'-l^2AR^2A'h_{1}'-l^2A^2Rh_{1}'R'-l^2A^2r_{1}R''+l^2A^2Rr_{1}''=0,
\end{align}
\begin{align}
	\frac{8Rr_{1}}{l^2}-\frac{2Rr_{1}}{A^2}+\frac{2g_{1}RR'}{A^2}+\frac{4h_{1}RA'R'}{A}-\frac{2r_{1}A'R'}{A}+Rh_{1}'R'\notag
	\\ +2h_{1}R'^2-\frac{2RA'r_{1}'}{A}-2R'r_{1}'+2h_{1}RR''-r_{1}R''-Rr_{1}''=0.
\label{firstordereomend}
\end{align}

Since the black hole is static it follows that the Killing vector $V = \frac{\partial}{\partial v}$ is hypersurface orthogonal, i.e. $V \wedge dV =0$. This implies
	\be
f_{1}(\rho)=c_{1},	
\ee
where $c_1$ is a constant. There is some residual gauge freedom related to rescaling $v$ and $r$. Using this gauge freedom the values the constant $c_{1}$ takes can be set to $0$ or $1$.\\
Furthermore the coordinate $v$ is only uniquely defined up to a translation by an arbitrary function of $\rho$. This makes us consider first order gauge transformations of the form
	\[
	v \rightarrow v + \alpha_1(\rho)
\]
	\[
	r  \rightarrow r + r^2 \,\alpha_2(\rho)
\]
	\be
	\rho \rightarrow \rho + r \, \alpha_3(\rho).
\ee
The reason we call these first order gauge transformations is that the zeroth order near horizon metric is unaffected by transformations of this kind and they only start showing up when going to first correction order. We have to be slightly careful though since the coordinate transformation will in general take us away from a gauge of the form (\ref{generalmetric}) by introducing $d\rho dr$ , $dr^2 $ and $dv dr$ terms. Observe though that $dr^2$ terms only enter at second correction order and thus we don't have to worry about them at this point. This means that there are $3$ arbitrary gauge functions $2$ of which get fixed by requiring that the transformation considered above keeps the metric of the form (\ref{generalmetric}). This leaves $1$ function to eliminate one of the $4$ undetermined functions entering the metric in (\ref{firstorder})
The particular gauge transformation we will consider is
	\[
	v = \widetilde{v} + \int^{\widetilde{\rho}} \frac{r_{1}(\rho_{1})}{A(\rho_{1})^2 R'(\rho_{1})}d\rho_{1},
\]
	\[
	r = \widetilde{r}+ \widetilde{r}^2 \; \frac{r_{1}(\widetilde{\rho})A'(\widetilde{\rho})}{R'(\widetilde{\rho})A(\widetilde{\rho})},
\]
	\be
	\label{firstordergauge}
	\rho = \widetilde{\rho} - \widetilde{r} \; \frac{r_{1}(\widetilde{\rho})}{R'(\widetilde{\rho})}.
\ee
This transformation sets $r_{1}(\rho)=0$. \\
Equations (\ref{firstordereomstart}) to (\ref{firstordereomend}) can than be solved algebraically by 
	\be
	f_{1}=c_1,
\ee
	\be
	g_{1}=-c_{1}\frac{R(R A'+A R')}{2 A R'^2},
\ee
	\be
	h_{1}=c_{1}\frac{R^2}{2 A^2 R'^2},
\ee
	\be
r_{1}=0.
\ee
We should pause here and consider for what coordinate ranges the first order corrections are small compared to the leading order. From \cite{Kaus:2009cg} we know for $\rho >> 1$ that $A(\rho)$ and $R(\rho)$ are proportional to $exp(\rho/l)$ and for $\rho << 1$ that $A = O(1)$ and $R=O(\rho^2)$. From this it follows that as long as $r<<1$ the first order corrections are small compared to the near horizon terms, i.e. there is no restriction on the range of $\rho$.

\subsection{Second order corrections}
Using the above results, expand the metric to next order in r.
	\[
	F(\rho,r)=-r^2+c_{1}r^3+f_{2}(\rho)r^4+O(r^5), \qquad G(\rho,r)=0\cdot r-c_{1}\frac{R(R A'+A R')}{2 A R'^2}r^2+g_{2}(\rho)r^3+O(r^4),
\]
	\be
	H(\rho,r)=1+c_{1}\frac{R^2}{2 A^2 R'^2}r+ h_{2}(\rho)r^2+O(r^3), \qquad I(\rho,r)=R(\rho)+0\cdot r +r_{2}(\rho)r^2+O(r^3).
\ee
It should be noted that at second or higher correction order there is no coordinate freedom left associated with a coordinate transformation equivalent to (\ref{firstordergauge}). This can be understood from the fact that the requirements $g_{rr}=g_{r\rho}=0$ and $g_{vr}=1$ fix all 3 gauge functions which arise or more geometrically that the original coordinate $v$ was defined uniquely up to a translation by an arbitrary function of $\rho$. However in the case $c_{1}=0$ there still exists gauge freedom associated with a transformation of the form $r \rightarrow c_{2} r$, $v\rightarrow \frac{v}{c_{2}}$. 

The same method as used in the previous section was applied to determine the equations of motion for the undetermined metric functions at second correction order, i.e. the Einstein equations were power series expanded in $r$ up to second order and the second order equations give rise to equations of motion for the metric functions to be determined. After some algebra these equations can be partially solved algebraically in terms of a single undetermined function
	\be
	f_{2}=\frac{7r_{2}}{3R}-\frac{2A^2r_{2}}{l^2R}+\frac{r_{2}A'^2}{R}-\frac{3c_{1}^2R^2}{8A^2R'^2}+\frac{2Ar_{2}A'R'}{R^2}-\frac{2A^2r_{2}R'^2}{R^3}-\frac{A^2R'r_{2}'}{R^2},
\ee
	\be
	g_{2}=\frac{4A^2(2r_{2}R'+Rr_{2}')}{3R^2},
\ee
	\be
	h_{2}=-\frac{2r_{2}}{R}.
\ee
The function $r_{2}$ is then determined by the equation
\begin{align}
\label{secondorderode}
	\frac{3c_{1}^2R^7}{A}-\frac{6c_{1}^2AR^7}{l^2}+\frac{3c_{1}^2R^7A'^2}{A}+11c_{1}^2R^6A'R'+c_{1}^2AR^5R'^2-44A^3R^2r_{2}R'^4+72A^5R^2r_{2}R'^4 \notag
	\\ -20A^3R^2r_{2}A'^2R'^4-80A^4Rr_{2}A'R'^5+24A^5r_{2}R'^6-8A^4R^2A'R'^4r_{2}'-4A^5R^2R'^4r_{2}''=0.
\end{align}
This equation could not be solved analytically. However it was solved numerically and the numeric results are shown in a later section. Following the approach applied in \cite{Kaus:2009cg} some analytic progress can be made though. It was argued that since the horizon of the brane world black hole is compact, $R(\rho)$ must vanish at some point which was chosen to be $\rho=0$ without loss of generality. Smoothness at $\rho=0$ then implies that in the neighbourhood of that point $A(\rho)$ and $R(\rho)$ have Taylor expansions of the form (\ref{Ataylor}) and (\ref{Rtaylor}).
Similarly smoothness at $\rho=0$ implies that the Taylor expansion of $r_{2}$ around that point is given by odd powers of $\rho$. Plugging this into the equation above determines $r_2$ to be given by
	\be
	\label{r2series}
r_{2}=-\frac{A_1}{5A_0^2}\rho^3+\Big(\frac{c_{1}^2}{56}+\frac{A_1A_0^2}{5l^2}+\frac{A_1}{70}\Big)\frac{\rho^5}{A_{0}^4}+\Big(\frac{2250c_{1}^2}{A_{0}^4l^2}-\frac{75c_{1}^2}{A_{0}^6}-\frac{58744A_1}{5A_0^2l^4}-\frac{552A_1}{A_{0}^4l^2}+\frac{1114A_1}{5A_{0}^6}\Big)\frac{\rho^7}{75600}+\ldots
\ee

	In contrast to $c_{1}$, $A_{1}$ is generally not gauge, but a free parameter like $A_{0}$. An exception to this occurs for $c_{1}=0$ in which case the coordinate transformation $r\rightarrow \frac{r}{\sqrt{\left|A_{1}\right|}}$, $v\rightarrow v\sqrt{\left|A_{1}\right|}$ sets the allowed values for $A_{1}$ to 0 or $\pm1$. The normalisation used here implies that $f_2(0)=A_1$.
	One should contrast the appearance of new free parameters to $4d$ extremal Reissner-Nordstr\"om. The respective expansion of the extremal Reissner-Nordstr\"om metric
	\be
	\label{RN}
	ds^2 = \Big(\frac{r^2}{Q^2}-\frac{2r^3}{Q^3}+\frac{3r^4}{Q^4}+\ldots \Big)dv^2+2dv dr + \Big(Q^2+2Q r + r^2+\ldots \Big)d\Omega^2,
\ee
	\be
	\label{RNgaugefield}
	A_{maxwell}^{RN}= \Big(-\frac{r}{Q}+\frac{r^2}{Q^2}-\frac{r^3}{Q^3}+\ldots \Big)dt,
\ee
only involves the single free parameter $Q$. The interpretation of this is that in the 4d case spherical symmetry fixes the corrections uniquely, whereas for the 5d bulk axisymmetry does not. We immediately conclude that, for the brane world black hole to be consistent with 4d General Relativity, the Israel matching conditions must fix any free parameter that appears in the correction terms. We will see that this is indeed the case in the next section. 

\sect{brane}

Following \cite{Kaus:2009cg}, we take the action of the brane to be
	\be
	S_{brane}=\int d^4z \sqrt{-h} \Big(-\sigma -\frac{1}{16\pi G_{4}}F_{ij}F^{ij} \Big),
\ee
where $\sigma$ is the brane tension, $G_{4}$ is the induced Newton constant and $F$ is the Maxwell field on the brane. We set $\sigma$ to its Randall-Sundrum value $\sigma=\frac{3}{4\pi G_{5} l}$, so that the induced cosmological constant on the brane vanishes. The induced Newton constant is then accordingly $G_{4}=\frac{G_{5}}{l}$. Since we are looking for a spherically symmetric brane world black hole, it is natural to assume the Maxwell field to be spherically symmetric as well. A duality transformation then allows us to take it purely electric so that 
	\be
	\star_{4}F=Q d\Omega,
\ee
where Q is the electric charge
	\be
	Q=\frac{1}{4\pi}\int_{S_{2}}\star F.
\ee
For this setup the Israel matching conditions are given by
	\be
	K_{ij}=\frac{1}{l}h_{ij}+l\Big( F_{i}^{k} F_{jk}-\frac{1}{4}h_{ij} F_{kl}F^{kl}  \Big).
	\label{Israel}
\ee
Here $K_{ij}$ is the extrinsic curvature and $h_{ij}$ is the induced metric on the brane. The most general brane position compatible with staticity and spherical symmetry is given by 
\be
\rho = \rho_{0}+r\,\rho_{1}+r^2\,\rho_{2}+\ldots.
\ee

The zeroth order solution $ds^2=A(\rho_{0})^2d\Sigma^2+R(\rho_{0})^2 d\Omega^2$ was determined in \cite{Kaus:2009cg} with the help of the zeroth order Israel equations
	\be
	\label{zerothisrael}
	\frac{A'(\rho_{0})}{A(\rho_{0})}=\frac{1}{l}-\frac{lQ^2}{2R(\rho_0)^4} \qquad ,	\frac{R'(\rho_{0})}{R(\rho_{0})}=\frac{1}{l}+\frac{lQ^2}{2R(\rho_0)^4}.
\ee
Here we will go beyond the zeroth order and investigate whether the bulk solution can be sliced by a brane of the above form at first and second correction order and determine the induced metric on the brane in case the bulk can be sliced. We then compare the induced solution on the brane to extremal Reissner Nordstr\"om, which it should assymptote in the limit where the ratio $Q/l$ is large, corresponding to large black holes.

		It should be pointed out that even though the near-horizon solution could be sliced by a brane, it is far from obvious on a mathematical level why the same should hold for the metric including corrections. Mathematically the main reason one expected to be able to slice the near horizon geometry with a brane is the $AdS_{2}$ symmetry factor, which appears in the near horizon limit. This symmetry enhancement resulted in the number of constraint equations and the number of free parameters being equal. However, when going beyond the near horizon limit, the $AdS_{2}$ symmetry is broken and thus, at every correction order, the number of constraint equations one expects to be independent is always strictly bigger than the number of free parameters at that order. Thus apriori one would not expect the near-horizon solution including correction terms to allow for a brane slicing. Physically however we do expect the full solution to exist, which of course implies that the Israel matching conditions can be fulfilled at each order and we will see below that this expectation is supported by our findings.
	
\subsection{First order corrections}

Expanding (\ref{Israel}) as a power series expansion in $r$ around $r=0$ gives rise to a set of constraint equations at each order. The leading order equations are the constraints (\ref{zerothisrael}). At first correction order, the Israel boundary equations give rise to 3 potentially independent constraints, which are given by

\begin{align}
0=&\rho_{1}-\frac{c_{1}A(\rho_{0})^2}{l}	+\frac{c_{1}l Q^2 A(\rho_{0})^2}{2R(\rho_0)^4}+c_{1}A(\rho_{0})A'(\rho_{0})+\frac{2\rho_{1}A(\rho_{0})A'(\rho_{0})}{l}-\rho_{1}A'(\rho_{0})^2 \notag\\
&-\frac{l Q^2\rho_{1}A(\rho_{0})A'(\rho_{0})}{R(\rho_0)^4} -\frac{c_{1}R(\rho_0)}{R'(\rho_0)}+\frac{2\rho_{1} l Q^2 A(\rho_{0})^2 R'(\rho_0) }{R(\rho_0)^5}-\rho_{1}A(\rho_{0})A''(\rho_{0}),\\
0=&\rho_{1}+\frac{2\rho_{1}A(\rho_{0})A'(\rho_{0})}{l}-\frac{l Q^2\rho_{1}A(\rho_{0})A'(\rho_{0})}{R(\rho_0)^4}-\frac{c_{1}l Q^2}{4 R(\rho_0)^2 R'(\rho_0)^2}-\rho_{1}A'(\rho_{0})^2-\frac{c_{1}R(\rho_0)}{R'(\rho_0)} \notag\\
&+\frac{c_{1}R(\rho_0)^2}{2l R'(\rho_0)^2}-\frac{c_{1}R(\rho_0)^2 A'(\rho_{0})^2}{2A(\rho_{0})R'(\rho_0)^2}+\frac{2\rho_{1}l Q^2A(\rho_{0})^2)R'(\rho_0)^2}{R(\rho_0)^5}-\rho_{1}A(\rho_{0})A''(\rho_{0}),\\
0=&-\frac{c_{1}R(\rho_0)^3}{2 A(\rho_0)^2 R'(\rho_0)}+\frac{\rho_{1}l Q^2 R'(\rho_0)}{R(\rho_0)^3}-\frac{2\rho_{1}R(\rho_0)R'(\rho_0)}{L}+ \rho_{1} R'(\rho_0)^2 +\rho_{1}R(\rho_0)R''(\rho_0).
\end{align}
The only undetermined parameter in these equations is $\rho_{1}$. However, after some algebra involving the equations of motion and (\ref{zerothisrael}), the above equations all simplify to the same equation
\begin{align}
\rho_{1}=\frac{c_{1}l^3A(\rho_{0})}{2(2A(\rho_{0})-l A'(\rho_{0}))(l^2+2(A(\rho_{0})-l A'(\rho_{0}))^2)}.
\end{align}
Thus, at first correction order, the bulk metric can indeed be sliced by a brane. It should be pointed out that the first order corrections are special in the sense that, in contrast to higher correction orders, all metric terms could be determined analytically in terms of the zeroth order solution. Therefore the question whether the bulk can be sliced by a brane at first correction order could also be answered affirmatively without involving any numerical methods beyond those which were needed to show that the near horizon limit can be sliced by a brane. As can be seen in the next subsection, this is different for higher order corrections since they involve functions which still need to be determined numerically.

\subsection{Second order corrections}

Applying the same method as discussed in the previous subsection in order to determine the constraint equations at second correction order, we obtain 4 potentially independent constraints arising from (\ref{Israel}). Their exact form can be found in the appendix. There are two undetermined parameters which enter these equations: $\rho_{2}$ and $A_1$. Using one of the constraint equations, it is possible to solve for $\rho_{2}$
	\be
	\rho_{2}=\frac{c_1 l^5 A(\rho_{0}) \Big( 4  A(\rho_{0})^2 -2 (4+c_1)l  A(\rho_{0})  A'(\rho_{0}) +l^2 (2+(4+c_1) A'(\rho_{0})^2) \Big) }{8(-2 A(\rho_{0})+l A'(\rho_{0}))^3(l^2+2( A(\rho_{0})-l A'(\rho_{0}))^2)^2},
\ee
where $Q$ was eliminated from the equation with the help of (\ref{zerothisrael}).

	As was the case for the first order corrections, the remaining 3 equations are independent at the level of the equations of motion, i.e. they all give different solutions as long as only equations (\ref{Aeom}),(\ref{Reom}),(\ref{firstordereomstart}) to (\ref{firstordereomend}) and the respective second correction order equations of motion are applied to simplify the constarints. However, when including all lower order Israel constraints and setting $c_1=0$, all $3$ equations simplify to 
	\be
	\label{casec0}
	0=-5\,l\, r_{2}(\rho_{0}) A'(\rho_{0})+ A(\rho_{0})\Big(6r_{2}(\rho_{0}) + l \,r_{2}'(\rho_{0})  \Big).
\ee
Doing the same in the case $c_1=1$, they all reduce to 

\begin{align}
\label{casec1}
0=&l^4 R(\rho_{0}) \Big(60 A(\rho_{0})^4-186 l A(\rho_{0})^3 A'(\rho_{0})-l^3 A(\rho_{0}) A'(\rho_{0}) \left(11+100 A'(\rho_{0})^2\right)\notag \\
&+l^2 A(\rho_{0})^2 \left(10+209 A'(\rho_{0})^2\right)+l^4 \left(-1+2A'(\rho_{0})^2+17 A'(\rho_{0})^4\right)\Big)+4 \left(-2 A(\rho_{0})+l A'(\rho_{0})\right)^3\notag \\
& \left(l^2+2 \left(A(\rho_{0})-l A'(\rho_{0})\right)^2\right)^2 \left(-5 l r_2(\rho_{0}) A'(\rho_{0})+A(\rho_{0}) \left(6r_2(\rho_{0})+l r_2'(\rho_{0})\right)\right).
\end{align}
Remember that the second free paramater $A_1$ enters the above equations through $r_{2}(\rho)$. This shows that in these two cases a brane slicing might be possible. However, in contrast to the first order corrections, here we have to refer to numerical methods to determine whether there is an $A_{1}$ solving the respective equation above. We will be doing this in the next section.

\sect{Numerical solutions}

Equation (\ref{secondorderode}) determines $r_2(\rho)$. It is solved numerically using the same strategy as in \cite{Kaus:2009cg}, i.e. we fix $A_0$ and $A_1$ and integrate (\ref{secondorderode}) with the help of the series expansion (\ref{r2series}) which fixes the inital data at $\rho=0$. Once $r_2(\rho)$ is determined it will be established numerically whether an $A_1$ exists, which satisfies (\ref{casec0}) or (\ref{casec1}) respectively, i.e. whether the bulk solution can be sliced by a brane. This is done by keeping $A_0$ fixed and varying $A_1$.

\begin{figure}[h]
\centerline{\includegraphics[width=.45\textwidth]{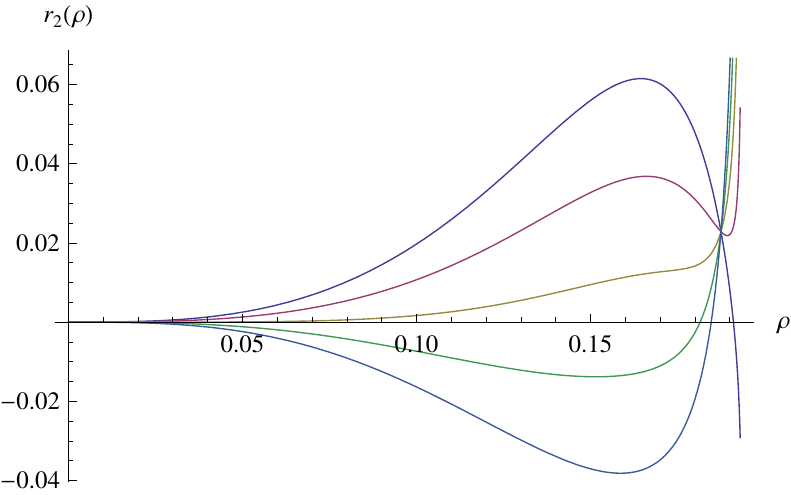}
\hspace{1cm}\includegraphics[width=.45\textwidth]{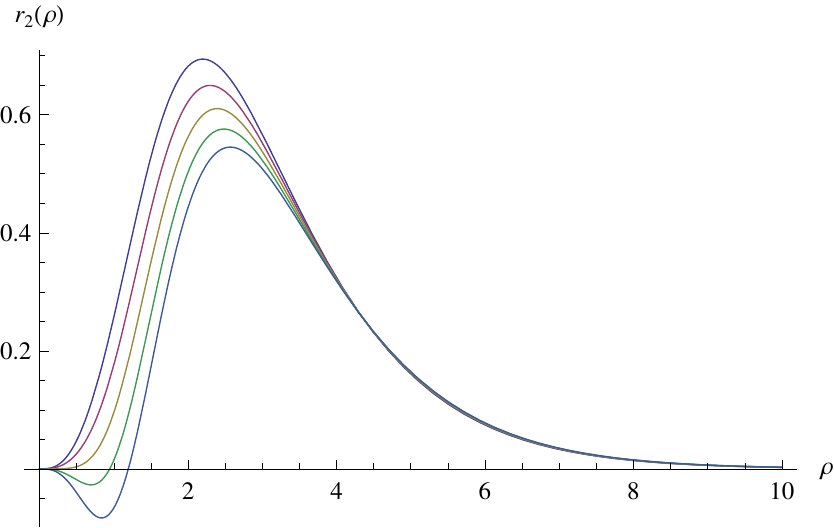}}
\caption{left: $A_0/l=0.1$, right: $A_0/l=0.65$, in each plot from top to bottom: $A_1=-1,-0.5,0,0.5,1$}
\label{fig1}
\end{figure}

	The general qualitative behaviour of $r_2(\rho)$ for different values of $A_0$ and $A_1$ in the case $c_1=1$ is shown in Fig. \ref{fig1}. It was determined in \cite{Kaus:2009cg} that for $A_0<l/2$, $R(\rho)$ diverges at a finite value of $\rho$. Here we find that for $A_0<l/2$ $r_2(\rho)$ diverges at the same value of $\rho$ independent of $A_1$. Furthermore that point coincides with the place at which the square of the near horizon Riemann tensor develops a curvature singularity. For $A_0>l/2$ , $r_2(\rho)$ converges to 0 for all $R_0$. The behaviour of $r_2(\rho)$ for small $\rho$ is given in (\ref{r2series}). In the case of $A_0>l/2$ solving (\ref{secondorderode}) for the leading order behaviour as $\rho\rightarrow \infty$ one finds that $r_2(\rho)$ is proportional to $\rho \, exp(-\rho/l)$. This implies that for $A_0>l/2$ the up to second order corrected bulk metric is asymptotically $AdS_5$ as $\rho\rightarrow \infty$.
	
\begin{figure}[h]
\centerline{\includegraphics[width=.45\textwidth]{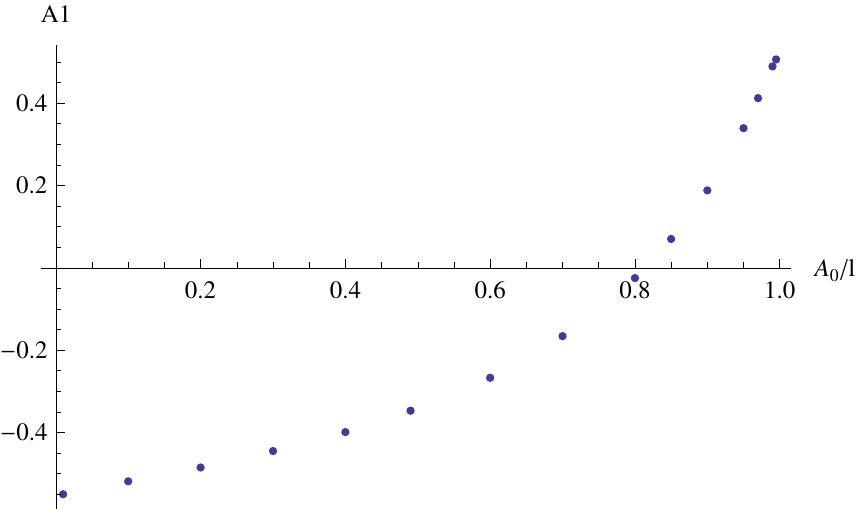}}
\caption{dependence of $A_1$ on $A_0/l$: there is $1 - 1$ relation between values of $A_0$ and $A_1$}
\label{fig2}
\end{figure}

	To determine whether the bulk including corrections can be sliced by a brane we first note that it was shown in \cite{Kaus:2009cg} that the near horizon bulk only allowed a brane slicing in the range $0< A_0 < l$. Thus we can limit ourselves to this range from now on. Furthermore we have seen that at first correction order a brane slicing is always possible assuming the near horizon solution allowed for it. At second order we have to distinguish the two cases $c_1=0$ and $c_1=1$. In the first case the matching condition which has to be fulfilled was determined above to be (\ref{casec0}). We find numerically that the only case which fulfills this is $A_1=0$, which corresponds to the trivial case of the original near horizon brane slicing. Thus we will from now on assume that $c_1=1$. In that case the Israel matching condition is (\ref{casec1}). It was determined numerically that there exists a unique $A_1$ fulfilling this equation given any $A_0$ in the range $0< A_0 < l$. This means that the second order corrected bulk can indeed be sliced by a brane. Fig. \ref{fig2} shows the dependence of $A_1$ on $A_0$. It should be mentioned that the zeroth order brane position slices the bulk in such a way that the spacetime containing the brane does not include the curvature singularity mentioned earlier in the case $0< A_0 < l/2$. This is still the case when including the correction terms considered here. Furhtermore given the range of $A_1$ in Fig. \ref{fig2} and the form of the second order metric corrections it follows that as long as $r<<1$, the correction terms are small compared to the zeroth order terms over the whole range of $\rho$. 
	
We now turn to analyze the brane world metric. Projecting onto the brane the induced metric is given to second correction order by

	\be
	ds^2=\Big(g_{vv}^{0} + g_{vv}^{1} + g_{vv}^{2} + ... \Big)r^2 dv^2 +\Big(g_{vr}^{0} + g_{vr}^{1} + g_{vr}^{2}+... \Big)2 dv dr + \Big(g_{\theta\theta}^{0} + g_{\theta\theta}^{1} + g_{\theta\theta}^{2} + ... \Big)d\Omega^2.
\ee
Where we have
\be 
g_{vv}^{0}=-A(\rho_{0})^2 r^2
\ee

\be 
g_{vv}^{1}=\Big(A(\rho_{0})^2-2\rho_{1}A(\rho_{0})A'(\rho_{0})\Big)r^3
\ee

\begin{align} 
g_{vv}^{2}= &\Big(\frac{7A(\rho_{0})^2 r_2(\rho_{0})}{3 R(\rho_{0})}-\frac{2A(\rho_{0})^4r_2(\rho_{0})}{l^2R(\rho_{0})}-\rho_{1}^2A'(\rho_{0})^2+2\rho_{1}A(\rho_{0})A'(\rho_{0})\notag \\
&-2\rho_{2}A(\rho_{0})A'(\rho_{0})+\frac{A(\rho_{0})^2r_2(\rho_{0})A'(\rho_{0})^2}{R(\rho_{0})}+\frac{2A(\rho_{0})^3r_2(\rho_{0})A'(\rho_{0})R'(\rho_{0})}{R(\rho_{0})^2}\notag \\
&-\frac{3R(\rho_{0})^2}{8R'(\rho_{0})^2}-\frac{2A(\rho_{0})^4r_2(\rho_{0})R'(\rho_{0})^2}{R(\rho_{0})^3}-\frac{A(\rho_{0})^4R'(\rho_{0})r_2'(\rho_{0})}{R(\rho_{0})^2}-\rho_{1}^2A(\rho_{0})A''(\rho_{0})\Big)r^4
\end{align}

\be 
g_{vr}^{0}=A(\rho_{0})^2
\ee
\be 
g_{vr}^{1}=2\rho_{1}A(\rho_{0})A'(\rho_{0})r
\ee
\be 
g_{vr}^{2}=\Big(\frac{\rho_{1}^2}{2}+2 \rho_{2}A(\rho_{0})A'(\rho_{0})+\rho_{1}^2 A'(\rho_{0})^2 -\frac{\rho_{1}R(\rho_{0})^2A'(\rho_{0})}{2R'(\rho_{0})^2A(\rho_{0})}-\frac{\rho_{1}R(\rho_{0})}{R'(\rho_{0})}+\rho_{1}^2 A(\rho_{0})A''(\rho_{0}) \Big)r^2
\ee
\be 
g_{\theta\theta}^{0}=R(\rho_{0})^2 
\ee
\be 
g_{\theta\theta}^{1}= 2 \rho_{1}R'(\rho_{0})R(\rho_{0})r
\ee

\be 
g_{\theta\theta}^{2}= \Big( 2 R(\rho_{0})r_2(\rho_{0})+2\rho_{2}R(\rho_{0})R'(\rho_{0})+\rho_{1}^2R'(\rho_{0})^2+\rho_{1}^2R(\rho_{0})R''(\rho_{0}) \Big)r^2
\ee

\begin{figure}[h]
\centerline{\includegraphics[width=.45\textwidth]{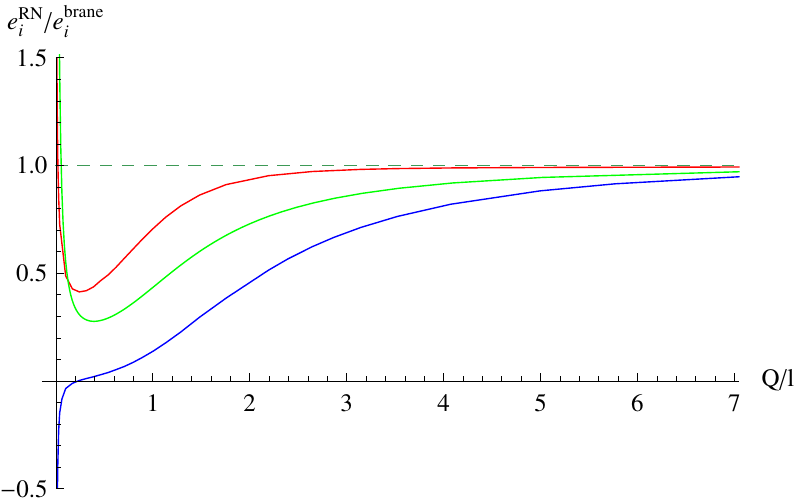}}
\caption{red $=e_1^{RN}/e_1^{brane}$, blue $= e_2^{RN}/e_2^{brane}$, green $= e_3^{RN}/e_3^{brane}$, all ratios converge to $1$ for large $Q/l$, i.e. the brane world black hole asymptotes $4d$ GR in that limit}
\label{fig3}
\end{figure}

The intention is to compare this to (\ref{RN}), since for large black holes we expect the induced black hole on the brane to asymptote the extremal Reissner Nordstr\"om solution. In order to investigate this we change the gauge of the above metric slightly to Eddington Finkelstein type coordinates by applying a transformation of the form $r\rightarrow f(r)$. So that the form of the metric becomes
	\be
	ds^2=\Big( f_{\textit{0th order}}+f_{\textit{1st order}} +f_{\textit{2nd order}} + ... \Big)r^2 dv^2 +2 dv dr + (g_{\textit{0th order}}+g_{\textit{1st order}}+g_{\textit{2nd order}}+ ...)d\Omega^2.
\ee
Given the metric in this form 3 quantities independent of the remaining gauge freedom, i.e. transformations of the form $r\rightarrow \epsilon \, r$ and $v\rightarrow v/\epsilon$, can be constructed out of the first and second order metric correction terms. 
the particular combination of 3 gauge independent terms we consider here is:
	\be
	e_{1}=\frac{f_{\textit{1st order}} }{g_{\textit{1st order}}} , \qquad e_{2}=\frac{f_{\textit{2nd order}}}{g_{\textit{2nd order}}} ,  \qquad  e_{3}=\frac{f_{\textit{1st order}}^2}{f_{\textit{2nd order}}}.
\ee
In the Reissner Nordstr\"om case these 3 quantities are given by $e_{1}^{RN}=\frac{1}{Q^4}$, $e_{2}^{RN}=-\frac{3}{Q^4}$ and $e_3^{RN}=-\frac{4}{3Q^2}$.
Having determined these quantities for both the brane world metric and the extremal Reissner Nordstr\"om solution, we calculate the respective ratios. Fig. \ref{fig3} shows the dependence of these quantities on the charge $Q$. \\

\begin{figure}[h]
\centerline{\includegraphics[width=.45\textwidth]{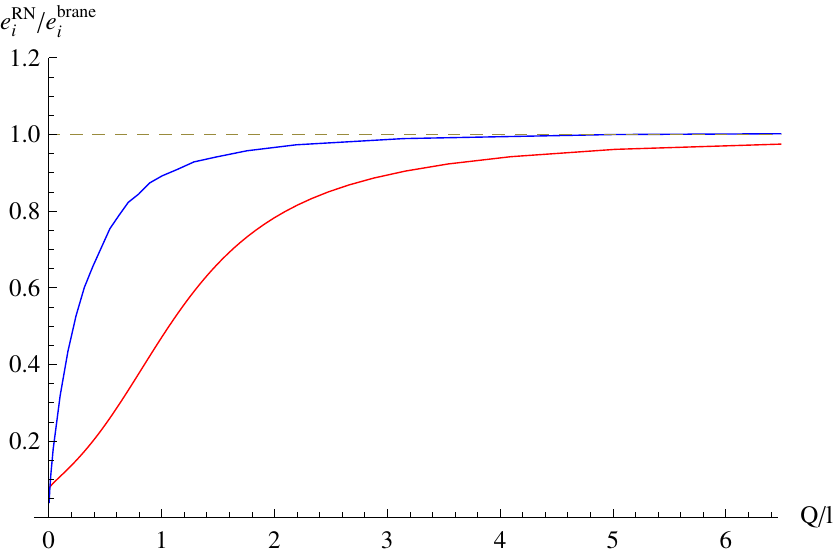}}
\caption{blue $=e_4^{RN}/e_4^{brane}$, red $= e_5^{RN}/e_5^{brane}$, all ratios converge to $1$ for large $Q/l$, i.e. the brane world black hole asymptotes $4d$ GR in that limit}
\label{fig4}
\end{figure}

As expected all ratios converge to $1$ for large charge over $AdS$ scale ratio. The next step is to compare the gauge fields of the two solutions. Expanding the gauge field of the brane world black hole as a power series in $r$ about $r=0$ in the same way as was done for the extremal Reissner Nordstr\"om solution in (\ref{RNgaugefield}), 2 more quantities independent of the gauge freedom $r\rightarrow \epsilon \, r$ and $v\rightarrow v/\epsilon$ can be obtained from the first and second order correction terms. In the gauge constructed above the power series expansion of the brane world black hole is given by
\be 
A_{maxwell}^{brane}=\Big(A_\textit{0th order}^{brane} + A_\textit{1st order}^{brane} + A_\textit{2nd order}^{brane} +... \Big) r dv
\ee
where
\be
A_\textit{0th order}^{brane}=-\frac{Q}{R(\rho_{0})^2}
\ee

\be
A_\textit{1st order}^{brane}=\frac{\rho_1 Q R'(\rho_{0})}{R(\rho_{0})^3 A(\rho_{0})^2}r
\ee

\be
A_\textit{2nd order}^{brane}=\Big(\frac{2 r_2(\rho_{0})}{R(\rho_{0})^3 A(\rho_{0})^4}+\frac{2\rho_2 R'(\rho_{0})}{R(\rho_{0})^3A(\rho_{0})^4}-\frac{2\rho_1^2  A'(\rho_{0}) R'(\rho_{0})}{R(\rho_{0})^3 A(\rho_{0})^5}-\frac{3\rho_1^2 R'(\rho_{0})^2}{R(\rho_{0})^4 A(\rho_{0})^4}+\frac{\rho_1^2 R''(\rho_{0})}{R(\rho_{0})^3 A(\rho_{0})^4}\Big)\frac{Q r^2}{3}.
\ee
\\
	From the gauge fields we construct the quantities
\be 
	e_4=\frac{A_\textit{1st order}^2}{A_\textit{2nd order}}
\ee

\be
	e_5=\frac{A_\textit{1st order}}{f_{\textit{1st order}}},
\ee
for both the extremal Reissner Nordstr\"om and the brane world solution. For the Reissner Nordstr\"om metric these are given by
$e_4^{RN}=-\frac{1}{Q}$ and $e_5^{RN}=-\frac{Q}{2}$. As done above we calculate the respective ratios between the $e_i^{RN}$ and the $e_i^{brane}$. Fig. \ref{fig4} depicts their dependence on the charge $Q$ and as before we have that the ratios tend to $1$ for large black holes.

\newpage
 This numerically extends the results of \cite{Kaus:2009cg} beyond the near horizon limit that the geometry of the brane world black hole asymptotes the corresponding 4d GR geometry for black holes large compared to the $AdS$ scale, i.e. when brane correction terms are expected to be subleading. However also analytically some progress can be made for large black holes. In \cite{Kaus:2009cg} it was determined that for large $Q$
 \be
 A(\rho_{0})=Q -\frac{3 l^2}{8 Q}+ O(\frac{1}{Q^3})
 \label{Aexpansion}
 \ee
 
 \be
 \label{Rexpansion}
 R(\rho_{0})=Q -\frac{ l^2}{8 Q}+ O(\frac{1}{Q^3}).
 \ee
 
 \be
 \rho_0=l \, Log(\frac{Q}{l}) + l \, Log(2) + ...
 \label{largerho0}
 \ee

Equations (\ref{zerothisrael}) then imply that for large $Q$

\be
 A'(\rho_{0})=\frac{Q}{L} -\frac{7 l}{8 Q}+ O(\frac{1}{Q^3})
\ee

\be
 R'(\rho_{0})=\frac{Q}{L} +\frac{3 l}{8 Q}+ O(\frac{1}{Q^3})
\ee
Furthermore from (\ref{secondorderode}) we know for large black holes, i.e. $Q >> 1$ or equivalently $\rho_0 >> 1$, that to leading order  $r_2(\rho_0)= C_0 \, \rho_0 \, exp(-\rho_0/l)$, where $C_0$ is some constant. It then follows from plugging this into (\ref{casec1}) that $C_0=1/4$. Making use of (\ref{largerho0}) it then implies that to leading order in $Q$

\be 
r_2(\rho_0) = \frac{l^2}{4} \frac{Log(Q/l)}{Q}+...
\ee
Using the above expansions it can similarly be determined that to leading order in Q
\be 
\rho_1= \frac{l}{2}+... \qquad \rho_2= \frac{l}{8}+...
\ee
Having obtained the leading order behaviour of all the building blocks neccessary we can now use those results to calculate the leading order behaviour of the $e_{i}^{brane}$, yielding

\be 
e_{1}^{brane}= \frac{1}{Q^4}+... \qquad e_{2}^{brane}=-\frac{3}{Q^4}+... \qquad e_3^{brane}=-\frac{4}{3Q^2}+...
\ee 

\be 
e_4^{brane}=-\frac{1}{Q}+... \qquad e_5^{brane}=-\frac{Q}{2}+...
\ee

Thus we have shown analytically up to second correction order that the geometry of the brane world black hole asymptotes the extremal Reissner Nordstr\"om solution for large black holes. 

\begin{figure}[h]
\centerline{\includegraphics[width=.45\textwidth]{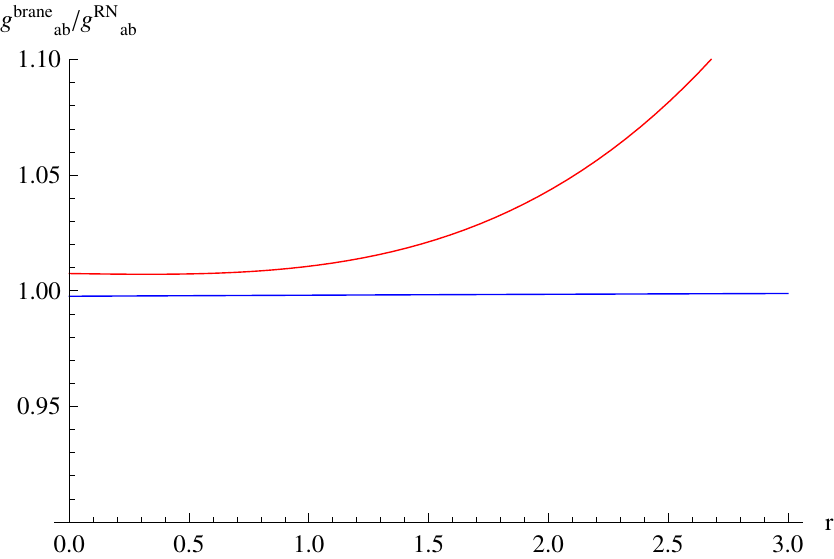}}
\caption{red $=g_{vv}^{brane}/g_{vv}^{RN}$, blue $= g_{\theta\theta}^{brane}/g_{\theta\theta}^{RN}$}
\label{fig5}
\end{figure}

	Assuming that the brane world black hole asymptotes $4d$ GR in the limit of large charge we can go back to investigate the convergence of the power series expansion, i.e. compare the series solution of the brane world black hole induced on the brane to the full solution of the extremal Reissner Nordstr\"om geometry as functions of $r$. Fig. \ref{fig5} shows the ratios  between the brane world black hole and the Reissner Nordstr\"om metric of the $g_{vv}$ and $g_{\theta\theta}$ components for a charge value of $Q/l = 10$. As can be seen the power series solution is a very good approximation to $4d$ GR up to $r=O(1)$. It follows from (\ref{Aexpansion}) and (\ref{Rexpansion}) that we expect the ratio of the  $g_{vv}$ components to asymptote 1 worse than the ratio of the $g_{\theta\theta}$ components as $r\rightarrow 0$. This behaviour is confirmed by the plot. Furthermore it should be pointed out that the $g_{\theta\theta}$ component of the brane world black hole approximates the full extremal Reissner Nordstr\"om solution very well even for $r > 1$. The reason for this is that the power series of the $g_{\theta\theta}$ component of the full solution terminates at second order. 
	
\sect{Discussion}

The aim of this paper was to extend the results of \cite{Kaus:2009cg}. The main loophole left open was that there is no guarantee that the near horizon geometry constructed in that paper extends to a full black hole solution and furthermore that if the near horizon limit can be extended that the full solution allows for a brane slicing. Even though a final proof of this could not be given here, strong evidence supporting that case was given: A perturbative expansion around the near horizon limit was set up and both the first and the second subleading correction terms were determined. It was shown that the metric including up to second order correction terms can still be sliced by a brane. This is remarkable as the inclusion of correction terms breaks the $AdS_{2}$ symmetry of the near horizon metric. That symmetry however was a main reason to expect the possibility of a brane slicing. Furthermore the induced metric on the brane was investigated and it was determined that the brane world black hole asymptotes $4d$ GR, i.e. the extremal Reissner Nordstrom metric, in the limit when the black hole is large compared to the $AdS$ scale.

	It should also be noted that the method to go beyond the near horizon limit which was applied here can in principle be extended to any correction order and that, while in practise more involved, conceptually the calculations carried out at every correction order $> 2$ are exactly the same as the calculations shown above for the second correction order: When solving the bulk metric 4 undetermined functions arise at each order 3 of which can be determined algebraically in terms of a single function. That single function is than determined by a second order ODE. Similarly there arise 4 possibly independent Israel matching conditions and 2 free parameters at every order. Even though the calculations were not included in this paper it was determined for the third order corrections that similarly to the second order corrections, these 4 equations reduce to 2 once lower order Israel constraints are included. Of course this does not prove anything about the general case.

	Given the results obtained in this paper the aim of any work following up from it is first and foremost to provide a proof that the full brane world black hole solution exists. The work presented here certainly suggests that this is the case. There are two immediate ways to procede given the results above. The first is to generalize this work by proving that a brane slicing is possible at arbitrary correction order. The main problem with this is that it seems that the Israel constraints at any given order can only be fulfilled assuming full knowledge of all lower order Israel conditions. A systematic way to do this could not be found so far. The second way in which progress could be possible is to determine the full bulk solution using numerical methods. For this the above results will be instrumental as well since with their help elliptic data of the respective boundary value problem can be determined. Of course the latter of the two options is preferable since it is constructive. 								
	
\bigskip

\begin{center}{\bf Acknowledgments}\end{center}

\medskip

\noindent I am very grateful to Harvey Reall for discussions and suggestions concerning the work presented here and beyond. I furthermore enjoyed discussions related to this work with Stephen Hawking and Ricardo Monteiro. 
 
\newpage
\sect{Appendix}
For completeness the second correction order Israel constraints are listed below in their original form

\begin{eqnarray}
0=&(2 \rho_{2}-\frac{7 \rho_{1} c_{1}}{2}+\frac{19 l Q^2 A(\rho_{0})^2 r_{2}(\rho_{0})}{6 R(\rho_{0})^5}-\frac{Q^2 A(\rho_{0})^4 r_{2}(\rho_{0})}{l R(\rho_{0})^5}-\frac{7 A(\rho_{0})^2 r_{2}(\rho_{0})}{3 l R(\rho_{0})}+\frac{2 A(\rho_{0})^4 r_{2}(\rho_{0})}{l^3 R(\rho_{0})}\notag \\
&+\frac{2 \rho_{2} A(\rho_{0}) A'(\rho_{0})}{l}-\frac{2 \rho_{1}c_{1} A(\rho_{0}) A'(\rho_{0})}{l}-\frac{\rho_{2} l Q^2 A(\rho_{0}) A'(\rho_{0})}{R(\rho_{0})^4}+\frac{\rho_{1} c_{1} L Q^2 A(\rho_{0}) A'(\rho_{0})}{R(\rho_{0})^4}\notag \\
&+\frac{A(\rho_{0}) r_{2}(\rho_{0}) A'(\rho_{0})}{3 R(\rho_{0})}-\frac{4 A(\rho_{0})^3 r_{2}(\rho_{0}) A'(\rho_{0})}{l^2 R(\rho_{0})}+\rho_{1} c_{1} A'(\rho_{0})^2+\frac{\rho_{1}^2 A'(\rho_{0})^2}{l}-\frac{\rho_{1}^2 l Q^2 A'(\rho_{0})^2}{2 R(\rho_{0})^4}\notag \\
&+\frac{l Q^2 A(\rho_{0})^2 r_{2}(\rho_{0}) A'(\rho_{0})^2}{2 R(\rho_{0})^5}-\frac{A(\rho_{0})^2 r_{2}(\rho_{0}) A'(\rho_{0})^2}{l R(\rho_{0})}+\frac{A(\rho_{0})r_{2}(\rho_{0}) A'(\rho_{0})^3}{R(\rho_{0})}+\frac{c_{1}^2 R(\rho_{0})^3}{2 A(\rho_{0})^2 R'(\rho_{0})^3}\notag \\
&-\frac{3 c_{1}^2 l Q^2}{16 R(\rho_{0})^2 R'(\rho_{0})^2}+\frac{3 c_{1}^2 R(\rho_{0})^2}{8 l R'(\rho_{0})^2}+\frac{\rho_{1} c_{1} R(\rho_{0})^2}{2 A(\rho_{0})^2 R'(\rho_{0})^2}+\frac{c_{1}^2 R(\rho_{0})^2A'(\rho_{0})}{4 A(\rho_{0}) R'(\rho_{0})^2}+\frac{9 c_{1}^2 R(\rho_{0})}{8R'(\rho_{0})}\notag \\
&+\frac{2 \rho_{2} l Q^2 A(\rho_{0})^2 R'(\rho_{0})}{R(\rho_{0})^5}-\frac{2 \rho_{1} c_{1} l Q^2 A(\rho_{0})^2 R'(\rho_{0})}{R(\rho_{0})^5}
+\frac{3 A(\rho_{0})^2 r_{2}(\rho_{0}) R'(\rho_{0})}{2 R(\rho_{0})^2}+\frac{A(\rho_{0})^4 r_{2}(\rho_{0}) R'(\rho_{0})}{l^2 R(\rho_{0})^2}\notag \\
&+\frac{4 \rho_{1}^2 l Q^2A(\rho_{0}) A'(\rho_{0}) R'(\rho_{0})}{R(\rho_{0})^5}+\frac{l Q^2 A(\rho_{0})^3 r_{2}(\rho_{0}) A'(\rho_{0}) R'(\rho_{0})}{R(\rho_{0})^6}-\frac{2 A(\rho_{0})^3 r_{2}(\rho_{0}) A'(\rho_{0})R'(\rho_{0})}{l R(\rho_{0})^2}\notag\\
&+\frac{5 A(\rho_{0})^2 r_{2}(\rho_{0}) A'(\rho_{0})^2 R'(\rho_{0})}{2 R(\rho_{0})^2}-\frac{5 \rho_{1}^2 l Q^2 A(\rho_{0})^2 R'(\rho_{0})^2}{R(\rho_{0})^6}-\frac{l Q^2 A(\rho_{0})^4 r_{2}(\rho_{0}) R'(\rho_{0})^2}{R(\rho_{0})^7}\notag \\
&+\frac{2 A(\rho_{0})^4 r_{2}(\rho_{0}) R'(\rho_{0})^2}{l R(\rho_{0})^3}-\frac{6 A(\rho_{0})^3 r_{2}(\rho_{0}) A'(\rho_{0}) R'(\rho_{0})^2}{R(\rho_{0})^3}+\frac{3 A(\rho_{0})^4 r_{2}(\rho_{0}) R'(\rho_{0})^3}{R(\rho_{0})^4}\notag \\
&-\frac{A(\rho_{0})^4 r_{2}'(\rho_{0})}{l^2 R(\rho_{0})}+\frac{A(\rho_{0})^2A'(\rho_{0})^2 r_{2}'(\rho_{0})}{2 R(\rho_{0})}-\frac{l Q^2 A(\rho_{0})^4 R'(\rho_{0}) r_{2}'(\rho_{0})}{2 R(\rho_{0})^6}+\frac{A(\rho_{0})^4 R'(\rho_{0}) r_{2}'(\rho_{0})}{lR(\rho_{0})^2}\notag\\
&-\frac{A(\rho_{0})^3 A'(\rho_{0}) R'(\rho_{0}) r_{2}'(\rho_{0})}{R(\rho_{0})^2}-\rho_{2} A(\rho_{0}) A''(\rho_{0})+\rho_{1} c_{1} A(\rho_{0}) A''(\rho_{0})+\frac{\rho_{1}^2 A(\rho_{0}) A''(\rho_{0})}{l}\notag\\
&-\frac{\rho_{1}^2 l Q^2 A(\rho_{0}) A''(\rho_{0})}{2 R(\rho_{0})^4}-\frac{3}{2} \rho_{1}^2 A'(\rho_{0}) A''(\rho_{0})
+\frac{A(\rho_{0})^2 r_{2}(\rho_{0}) A'(\rho_{0})A''(\rho_{0})}{R(\rho_{0})}+\frac{\rho_{1}^2A'(\rho_{0})}{2 A(\rho_{0})}\notag\\
&+\frac{A(\rho_{0})^3 r_{2}(\rho_{0}) R'(\rho_{0}) A''(\rho_{0})}{R(\rho_{0})^2}+\frac{\rho_{1}^2 l Q^2 A(\rho_{0})^2R''(\rho_{0})}{R(\rho_{0})^5}+\frac{A(\rho_{0})^3 r_{2}(\rho_{0}) A'(\rho_{0}) R''(\rho_{0})}{R(\rho_{0})^2} \notag\\
&+\frac{3 c_{1}^2 R(\rho_{0})^2 R''(\rho_{0})}{8R'(\rho_{0})^3}+\frac{\rho_{1} c_{1} R(\rho_{0}) R''(\rho_{0})}{R'(\rho_{0})^2}-\frac{2 A(\rho_{0})^4 r_{2}(\rho_{0})R'(\rho_{0})R''(\rho_{0})}{R(\rho_{0})^3}+\frac{5 A(\rho_{0})^2 r_{2}'(\rho_{0})}{2 R(\rho_{0})} \notag \\
& -\rho_{2} A'(\rho_{0})^2-\frac{A(\rho_{0})^4r_{2}'(\rho_{0})R''(\rho_{0})}{2R(\rho_{0})^2}-\frac{A(\rho_{0})^4 R'(\rho_{0}) r_{2}''(\rho_{0})}{2 R(\rho_{0})^2}-\frac{1}{2} \rho_{1}^2 A(\rho_{0}) A'''(\rho_{0})
\end{eqnarray}

\begin{eqnarray}
0=&\frac{3 l Q^2 A(\rho_{0})^2 r_{2}(\rho_{0})}{R[a]^5}-\frac{2 A(\rho_{0})^2 r_{2}(\rho_{0})}{l R(\rho_{0})}-\frac{5 c_1 \rho _1}{2}-\frac{\rho _1^2}{2 l}+\frac{l Q^2 \rho _1^2}{4 R(\rho_{0})^4}+2 \rho_2 +\frac{\rho _1^2 A'(\rho_{0})}{A(\rho_{0})}+\frac{2 A(\rho_{0}) \rho _2 A'(\rho_{0})}{l}\notag\\
&-\frac{l Q^2 A(\rho_{0}) \rho _2 A'(\rho_{0})}{R(\rho_{0})^4}+\frac{\rho _1^2 A'(\rho_{0})^2}{l}-\frac{1}{2} A(\rho_{0}) \rho _1^2 A'''(\rho_{0})
-\frac{l Q^2 \rho _1^2 A'(\rho_{0})^2}{2 R(\rho_{0})^4}-\rho _2 A'(\rho_{0})^2+\frac{R(\rho_{0})^2 c_1 \rho _1}{A(\rho_{0})^2 R'(\rho_{0})^2}\notag\\
&+\frac{R(\rho_{0})^2 c_1 \rho _1 A'(\rho_{0})^2}{2 A(\rho_{0})^2
R'(\rho_{0})^2}+\frac{l Q^2 c_1 \rho _1}{2 R(\rho_{0})^3 R'(\rho_{0})}+\frac{R(\rho_{0}) c_1 \rho _1}{l R'(\rho_{0})}-\frac{R(\rho_{0}) c_1 \rho _1 A'(\rho_{0})}{A(\rho_{0}) R'(\rho_{0})}+\frac{4 A(\rho_{0})^2 r_{2}(\rho_{0})
R'(\rho_{0})}{R(\rho_{0})^2}\notag\\
&+\frac{2 l Q^2 A(\rho_{0})^2 \rho _2 R'(\rho_{0})}{R(\rho_{0})^5}+\frac{4 l Q^2 A(\rho_{0}) \rho _1^2 A'(\rho_{0}) R'(\rho_{0})}{R(\rho_{0})^5}-\frac{5 l Q^2 A(\rho_{0})^2 \rho _1^2 R'(\rho_{0})^2}{R(\rho_{0})^6}\notag\\
&+\frac{2 A(\rho_{0})^2 r_{2}'(\rho_{0})}{R(\rho_{0})}+\frac{A(\rho_{0}) \rho _1^2 A''(\rho_{0})}{l}-\frac{l Q^2 A(\rho_{0}) \rho _1^2 A''(\rho_{0})}{2 R(\rho_{0})^4}-A(\rho_{0}) \rho _2 A''(\rho_{0})\notag\\
&+\frac{R(\rho_{0})c_1 \rho _1 R''(\rho_{0})}{R'(\rho_{0})^2}-\frac{3}{2} \rho _1^2 A'(\rho_{0}) A''(\rho_{0})-\frac{R(\rho_{0})^2 c_1 \rho _1 A''(\rho_{0})}{2 A(\rho_{0}) R'(\rho_{0})^2}+\frac{l Q^2 A(\rho_{0})^2 \rho _1^2 R''(\rho_{0})}{R(\rho_{0})^5}\notag\\
&+\frac{l Q^2 c_1 \rho _1 R''(\rho_{0})}{2 R(\rho_{0})^2 R'(\rho_{0})^3}-\frac{R(\rho_{0})^2 c_1 \rho _1 R''(\rho_{0})}{l R'(\rho_{0})^3}+\frac{R(\rho_{0})^2 c_1 \rho _1 A'(\rho_{0}) R''(\rho_{0})}{A(\rho_{0}) R'(\rho_{0})^3}
\end{eqnarray}

\begin{eqnarray}
0=&-\frac{\rho _1^2}{l}+\frac{l Q^2 \rho _1^2}{2 R(\rho_{0})^4}-2 \rho _2+\frac{2 \rho _1^2 A'(\rho_{0})}{A(\rho_{0})}-\frac{c_1 R(\rho_{0})^2
\rho _1}{A(\rho_{0})^2 R'(\rho_{0})^2}
\end{eqnarray}

\begin{eqnarray}
0=&-\frac{3 c_1 R(\rho_{0})^2 \rho _1}{2 A(\rho_{0})^2}+\frac{l Q^2 r_2(\rho_{0})}{R(\rho_{0})^3}-\frac{2 R(\rho_{0}) r_2(\rho_{0})}{l}+\frac{c_{1}^2 R(\rho_{0})^5}{4 A(\rho_{0})^4R'(\rho_{0})^3}+\frac{c_1 R(\rho_{0})^3 \rho _1 A'(\rho_{0})}{A(\rho_{0})^3 R'(\rho_{0})}-\frac{R(\rho_{0}) \rho _1^2 R'(\rho_{0})}{2 A(\rho_{0})^2}\notag\\
&+\frac{l Q^2 \rho _2 R'(\rho_{0})}{R(\rho_{0})^3}-\frac{2 R(\rho_{0})\rho_2 R'(\rho_{0})}{l}+3 r_2(\rho_{0}) R'(\rho_{0})-\frac{\rho _1^2 R'(\rho_{0})^2}{l}-\frac{3 l Q^2 \rho _1^2 R'(\rho_{0})^2}{2 R(\rho_{0})^4}+\rho _2 R'(\rho_{0})^2\notag\\
&+R(\rho_{0}) r_2'(\rho_{0})+\frac{l Q^2 \rho _1^2 R''(\rho_{0})}{2 R(\rho_{0})^3}-\frac{R(\rho_{0}) \rho _1^2 R''(\rho_{0})}{l}+R(\rho_{0}) \rho _2 R''(\rho_{0})+\frac{c_1 R(\rho_{0})^3 \rho _1 R''(\rho_{0})}{2 A(\rho_{0})^2R'(\rho_{0})^2}\notag\\
&+\frac{3}{2} \rho _1^2 R'(\rho_{0}) R''(\rho_{0})+\frac{1}{2} R(\rho_{0}) \rho _1^2 R'''(\rho_{0})
\end{eqnarray}

\end{document}